\documentclass[fleqn]{article}
\usepackage{latexsym}
\usepackage[latin1]{inputenc}
\usepackage{graphicx}
\newcommand{\T}{{\rm tr}}

\newcommand{\cH}{{\cal H}}

\newcommand{\0}{{\mathbf 0}}
\newcommand{\1}{{\mathbf 1}}
\newcommand{\rc}{{\rm anti}}

\def\bbbr{{\rm I\!R}}
\def\bbbc{{\mathchoice {\setbox0=\hbox{$\displaystyle
\rm C$}\hbox{\hbox
to0pt{\kern0.4\wd0\vrule height0.9\ht0\hss}\box0}}
{\setbox0=\hbox{$\textstyle\rm C$}\hbox{\hbox
to0pt{\kern0.4\wd0\vrule height0.9\ht0\hss}\box0}}
{\setbox0=\hbox{$\scriptstyle\rm C$}\hbox{\hbox
to0pt{\kern0.4\wd0\vrule height0.9\ht0\hss}\box0}}
{\setbox0=\hbox{$\scriptscriptstyle\rm C$}\hbox{\hbox
to0pt{\kern0.4\wd0\vrule height0.9\ht0\hss}\box0}}}}
\begin{document}
%%%%%%%%%%%%%%%%%%%%%%%%%%%%%%%%%%%%%%%%%%%%%%%%%%%%%%%%%%%%%%%%%%%%%%%%
\title{On Concurrence and Entanglement of Rank Two
Channels\footnote{Open Sys. \& Information Dyn. 12 (2005) 1-14.
An appendix is added. An error
in (\ref{line10}) - (\ref{line12})
and known (to me) misspellings are corrected.}}
  \author{Armin Uhlmann}
  \date{University of Leipzig,\\ Institute for Theoretical Physics}
  \maketitle

  %%%%%%%%%%%%%%%%%%%%%%%%%%%%%%%%%%%%%%%%%%%%%%%%%%%%%%%%%%%%%%%%%%%%%%%%
\begin{abstract}
      Concurrence and further entanglement quantifiers
      can be computed explicitly for channels of rank
      two if representable by just two Kraus operators.
      Almost all details are available
      for the subclass of rank two 1-qubit channels.
      There is a simple geometric picture beyond,
      explaining nicely the role of anti-linearity.
\end{abstract}
  %%%%%%%%%%%%%%%%%%%%%%%%%%%%%%%%%%%%%%%%%%%%%%%%%%%%%%%%%%%%%%%%%%%%%%%%

%\makeindex
\section{Introduction}

The aim of the present paper is to study completely
positive (i.~e. ``cp''--) maps $\Phi$ of rank two,
in particular, some of its entanglement properties.
These maps can be Kraus represented by
\begin{equation} \label{kraus1}
\Phi(X) = \sum_{j=1}^m  A_j X A_j^*
\end{equation}
with linear independent operators
\begin{equation} \label{kraus2}
A_j \, : \qquad \cH_d \mapsto \cH_2
\end{equation}
from an Hilbert space $\cH_d$ of dimension $d$ into
2-dimensional Hilbert space. The integer $m$ will be
called the {\em length} of $\Phi$. The complex linear
space generated by the Kraus operators (\ref{kraus2})
does not depend on the choice of the Kraus operators and
will be referred to as {\em Kraus space} of $\Phi$ and it
is denoted by Kraus$(\Phi)$. Its dimension is the
length of $\Phi$. These definitions are not bound to
the particular class of cp-maps satisfying
(\ref{kraus2}), to which the paper is devoted.
\medskip

$\Phi$ being of rank two, the output $\Phi(X)$ for
Hermitian $X$ enjoys only two independent unitary
invariants, the trace and the determinant.
In case of a quantum channel, i.~e.~a
trace preserving cp-map, only the determinant counts.
In the next section a remarkable and,
perhaps, not completely evident way to express
$\det \Phi(X)$ for pure input states is deduced.
\medskip

By the  important
papers of Hill and Wootters, \cite{HilWoo97},
and of Wootters, \cite{Woo97}, ``concurrence''
has been proved an important tool in the entanglement
problem (with respect to the partial trace). Its
conceptional roots go back to
the seminal work of Bennett et al, \cite{BenFucSmo96}.
See also the review \cite{Woo01} of Wootters.

The {\em concurrence,} $C(\Phi ; .)$, of $\Phi$ can be
defined generally as the solution of an optimization
task: It is the largest convex function on the input
state space, coinciding for every pure input state with
twice the square root of the output's
second symmetric function.
The second symmetric function of an operator on $\cH_2$
is its determinant. Thus, the concurrence is the largest
convex function on the input state space satisfying
\begin{displaymath}
C(\Phi ; \pi) = 2  \bigl( \det \Phi(\pi) \bigr)^{1/2},
\quad \pi \hbox{ pure.}
\end{displaymath}
The factor two does not play a decisive role and is for
historical reasons only. If it is neglected, one has just
to re-scale some constants.
It is sometimes useful to extend the definition to the
positive cone of the input system by requiring degree one
homogeneity, see section 3.

For most cp-maps an explicit expression for the concurrence
is unknown. Exceptions are the rank and length two
cases, as can be seen from \cite{Woo97} and \cite{Uh00c}.
Fortunately, based on \cite{Uh00c}, just
for these cases one can prove ``flatness'' of the convex roof
$C(\Phi; .)$ : If $\omega$ is an input state, there
are pure input states $\pi_1, \pi_2, \dots$ such that\\
a) $\omega$ is a convex combination of the $\pi_k$, and\\
b) $C(\Phi ; .)$ is {\em constant} on the convex set
generated by all the $\pi_k$.
\medskip

A rather complete picture can be given for 1-qubit channels
of length two. The linear structure of 1-qubit channels is
well studied in Ruskai et al, \cite{RSW00} and in
Verstraete and Verschelde, \cite{VV02}, following
Fujiwara and Algoet, \cite{FA98}. This line of thinking
is going back to Gorini and Sudarshan, \cite{GS76}, who
classified all affine maps of the d-dimensional ball into
itself.
However, if we need more than two Kraus operators
to represent a 1-qubit cp-map, then we mostly loose
the control on
the flatness of $C$ and of other entanglement measures.
Exceptions are some trivial cases in which $\det \Phi$
is constant on the set of all pure states.
\medskip

Let us now see, as an illustration, what happened with the
concurrence for a non-degenerate 1-qubit channel of length two:
The input Bloch space is covered by parallel straight lines on
which the concurrence is constant.  For every mixed input state
$\omega$ there is exactly one such line containing $\omega$. It
crosses the Bloch sphere at two pure input states, say $\pi_1$ and
$\pi_2$. The determinants of $\Phi(\pi_j)$, $j=1,2$, coincide.
They determine the value of the concurrence along the line in
question. Therefore, because of their parallelism, we have to know
just one of these lines to compute $C$. Fortunately, there is a
distinguished line on which $C$ is zero. To get that line we
have to find the two pure input states which are mapped onto pure
outputs by $\Phi$. That is, one has to solve the quadratic
equation $\det \Phi(\pi) = 0$.\\
An input vector $\tilde \psi$  will be called
{\em $\Phi$-separable} if there is an output
vector $\tilde \varphi$ such that
\begin{equation} \label{separabel1}
\Phi(|\tilde \psi \rangle\langle \tilde \psi|) =
|\tilde \varphi \rangle\langle \tilde \varphi| .
\end{equation}
{\em Let $\Phi$ be a non--degenerate 1-qubit
channel of length two. Then the Bloch--space is covered
by parallel lines of constant concurrence. Their geometry
is completely determined by the positions of the
$\Phi$-separable input vectors.}
\medskip

Let us return to the line of constant concurrence
containing a given $\omega$.
If we draw a plane through $\omega$ perpendicular to
that line, we may ask for the locus of
points with equal concurrence. The answer is an ellipse.
Thus, every plane perpendicular to a line of constant
concurrence is covered by ellipses of constant
concurrence:
{\em $C = {\rm constant}$ defines an ellipse--based
cylinder in Bloch--space.}

In the degenerate case, in which $\det \Phi(\pi) = 0$
has a double root, $C$ becomes linear (affine) along
planes.
\medskip

If the concurrence is flat, one can use almost literally
Wootters' reasoning in treating the
$(2 \times 2)$-entanglement of formation.
By the Stinespring dilatation theorem, every channel is
unitary equivalent to a partial trace, provided
the latter is
restricted to density operators with a suitably
selected support space. From this perspective it
becomes clear, how one has
to define the functional, which reproduce entanglement
of formation, \cite{BenFucSmo96}, according to the
Stinespring equivalence. This entanglement functional
will be denoted by $E(\Phi ; .)$. It is the largest
convex function on the input states satisfying
\begin{displaymath}
E(\Phi ; \pi) = S(\Phi(\pi)), \quad \pi \hbox{ pure}
\end{displaymath}
where $S$ denotes the von Neumann entropy.
Taking into account
what has been said above, one can write down analytic
expressions for $E(\Phi ; .)$ as a function of
$C(\Phi ; .)$ {\em for all quantum channels
of rank and length two.} Though the numerical values
of $C$ and $E$ are quite different in nature, their
geometry is isomorphic: They are constant along the
same straight lines of the input Bloch space.

\section{The determinant}

Let $\Phi$ as given by (\ref{kraus1}) and (\ref{kraus2}).
We look for $\det \Phi(X)$, rank$(X)=1$.
There are several ways to do so without insisting
to rank two, aiming at concurrences in general, see
Rungta et al, \cite{RBCHMW}, Albeverio and Fei, \cite{AF01},
and Mintert et al, \cite{MKB04}. Here we follow \cite{Uh01}
and \cite{Uh03c} in using anti-linear operators tailored
just to the rank two case.

Hilbert spaces of dimension two come with an exceptional
anti-unitary operator, the spin-flip $\theta_f$.
(The index ``f'' remembers Fermi and ``fermion''.)
We choose a reference basis, $|0\rangle$, $|1\rangle$,
and fix the phase according to
\begin{equation} \label{flip1}
\theta_f (c_0 |0\rangle + c_1 |1\rangle) =
c_1^* |0\rangle - c_0^* |1\rangle,
\end{equation}
or, in a self-explaining way, by
\begin{displaymath}
\theta_f \pmatrix{c_0 \cr c_1} =
\pmatrix{0 & 1 \cr -1 & 0}_{\rc} \pmatrix{c_0 \cr c_1}
= \pmatrix{c_1^* \cr -c_0^*} .
\end{displaymath}
($\to$ See also appendix,1.) We need the well known equation
\begin{equation} \label{flip2}
\theta_f Y^* \theta_f Y = - ( \det Y ) \, \1 .
\end{equation}

The anti-linear operator $A_j^* \theta_f A_k$ is well defined
for Kraus operators (\ref{kraus2}). It acts on $\cH_2$, and
its Hermitian part, $\vartheta_{jk}$, reads
\begin{equation} \label{anti1}
\vartheta_{jk} = {1 \over 2} \bigl(
A_j^* \theta_f A_k - A_k^* \theta_f A_j \bigr) .
\end{equation}
{\bf Theorem 1} \, {\it
Let $A_1, \dots, A_m$ denote the Kraus operators
(\ref{kraus2}) of a cp-map $\Phi$ of rank 2, and $\vartheta_{jk}$
defined according to (\ref{anti1}). Then
\begin{equation} \label{det2}
\det \Phi(|\psi_2 \rangle\langle \psi_1 |) =
\sum_{j < k}  \, \langle \psi_1 , \vartheta_{jk} \psi_1 \rangle  \,
\langle \psi_2 , \vartheta_{jk} \psi_2 \rangle^*,
\quad \psi_i \in \cH_d .
\end{equation}
The complex--linear span
of the operators $\vartheta_{jk}$ is uniquely
associated to $\Phi$.$\diamond$}

I use the ad hoc notation ``(first) derived Kraus--space'',
abbreviated Kraus'$(\Phi)$,
for the linear space generated by the operators
(\ref{anti1}). It is a linear space over the complex numbers
as $(c \vartheta)^*$ equals $\vartheta c^* = c \vartheta$
for Hermitian anti--linear operators.

To prove (\ref{det2}), we apply (\ref{flip2}) to
$Y = \Phi(X)$ and take the trace:
\begin{equation} \label{det1}
\det \Phi(X) = - {1 \over 2} \T \,
\sum_{jk} (A_k^* \theta_f A_j) X^* (A_j^* \theta_f A_k) X
\end{equation}
We insert $X = |\psi_1 \rangle\langle \psi_2|$ to obtain
\begin{displaymath}
\det \Phi(|\psi_1\rangle\langle \psi_2|) = - \sum_{j<k}
\langle \psi_2, (A_k^* \theta_f A_j) \psi_2 \rangle \cdot
\langle (A_j^* \theta_f A_k) \psi_1, \psi_1\rangle
\end{displaymath}
by respecting the anti--linearity rules. We observe
\begin{displaymath}
\langle \psi_2, A_k^* \theta_f A_j \psi_2 \rangle =
\langle A_k \psi_2,  \theta_f A_j \psi_2 \rangle = -
\langle A_j \psi_2,  \theta_f A_k \psi_2 \rangle .
\end{displaymath}
This tells us, that only the Hermitian parts of the operators
$A_j^* \theta_f A_k$ count, and we can replace them by
the operators (\ref{anti1}). Thus, (\ref{det1}) is proved.
Two elements of the Kraus space relate to (\ref{anti1}) as
\begin{equation} \label{trans1}
(\sum a_j A_j)^* \theta_f (\sum b_k A_k) -
(\sum b_k A_k)^* \theta_f (\sum a_j A_j) =
\sum_{jk} a_j^* b_k^* \vartheta_{jk},
\end{equation}
which proves the second assertion of the theorem.
\medskip

In changing to another set of Kraus operators for $\Phi$,
say $\tilde A_1, \tilde A_2, \dots$,
 the transformation coefficients form a
unitary matrix. Together with (\ref{trans1}) one obtains
\begin{equation} \label{trans1a}
\tilde A_k = \sum_j u_{jk} A_j, \quad
\tilde \vartheta_{mn} = \sum_{jk} u_{jm} u_{kn} \vartheta_{jk},
\end{equation}
with the indexed $\tilde \vartheta$ defined as in (\ref{anti1}).
By the help of (\ref{trans1a}) one gets
\begin{displaymath}
\sum \tilde \vartheta_{mn} X \tilde \vartheta_{mn} =
\sum u_{jm} u_{kn}   u_{rm}^* u_{sn}^*
\vartheta_{jk} X \vartheta_{rs} =
\sum \vartheta_{jk} X \vartheta_{jk} .
\end{displaymath}
These calculations show:\\
{\bf Lemma 1} \, {\it
The completely co-positive super-operator
\begin{equation} \label{det3}
\Phi'(X) := \sum_{j<k} \vartheta_{jk} X^* \vartheta_{jk}
\end{equation}
is uniquely associated to $\Phi$ and is called
``(first) derivative'' of $\Phi$.$\diamond$}

From (\ref{det2}) and (\ref{det3}) one concludes
\begin{equation} \label{det4}
\det \Phi(|\psi_2 \rangle\langle \psi_1|) =
\langle \psi_1, \Phi'( | \psi_1 \rangle\langle \psi_2 |)
\psi_2 \rangle
\end{equation}

\subsection{Length two}

Now let (\ref{kraus1}) be of length two and let us
denote the two Kraus operators in (\ref{kraus2}) by
$A$ and $B$. From them the anti-linear operator
$\vartheta$ is constructed according to (\ref{anti1}).
After choosing reference bases in the two Hilbert
spaces, we get matrix representations
\begin{equation} \label{matrix1}
A = \left( \begin{array}{cccc}
a_{00} & a_{01} & a_{02} & \dots \cr
a_{10} & a_{11} & a_{12} & \dots
\end{array} \right),
\quad
B = \left( \begin{array}{cccc}
b_{00} & b_{01} & b_{12} & \dots  \cr
b_{10} & b_{11} & b_{12} & \dots
\end{array} \right) .
\end{equation}
$A^* \theta_f B$ acts anti-linearly on $\cH_d$ with matrix
entries
\begin{displaymath}
\{ A^* \theta_f B \}_{mn} =
( a_{0m} b_{1n} - a_{1m} b_{0n})^*
\end{displaymath}
in the chosen basis. ($\to$ See also appendix,1.)
The matrix of an Hermitian anti-linear
operator is symmetric in every basis. Hence, we get for
the matrix entries of $\vartheta$
\begin{equation} \label{matrix2}
\{ \vartheta \}_{mm} =
(a_{0m} b_{1m} + a_{0m} b_{1m})^* \; ,
\end{equation}
\begin{displaymath}
\{ \vartheta \}_{mn} =
{1 \over 2} (a_{0m} b_{1n} + a_{0n} b_{1m}
- a_{1m} b_{0n}  - a_{1n} b_{0m})^*, \quad m \neq n \; .
\end{displaymath}

\noindent \underline{1-qubit channels of length two}
 can be given by
\begin{equation} \label{l2r2.1}
A = \pmatrix{a_{00} & 0 \cr 0 & a_{11}}, \quad
B = \pmatrix{0 & b_{01} \cr b_{10} & 0} ,
\end{equation}
up to unitary equivalence, \cite{RSW00}.
To get trace preserving, one needs restrictions.
But we do not need them. ($\to$ See also appendix,2.)
Just by inserting into
(\ref{matrix2}), $\vartheta$ appears to be
\begin{equation} \label{l2r2.2}
\vartheta = \pmatrix{z_0^2 & 0 \cr 0 & - z_1^2}_{\rc}, \quad
z_0^2 = (b_{10} a_{00})^*, \quad z_1^2 = (b_{01} a_{11})^*
\end{equation}
and (\ref{det2}) results in
\begin{equation} \label{l2r2.3}
\det \Phi( \pmatrix{a_0 a_0^* & a_0 a_1^* \cr a_1 a_0^* & a_1 a_1^*} )
= | (z_0 a_0^* + z_1 a_1^*) \, (z_0 a_0^* - z_1 a_1^*) |^2 .
\end{equation}
The map $\Phi$ is called {\em non-degenerate}
if $z_0 z_1 \neq 0$. Then there are two linear independent
$\Phi$-separable input vectors.
%We arrive at the case sketched in the introduction.

If $\Phi$ is \underline{degenerate}, there are several
cases: Either one
of the numbers $z_0, z_1$ is zero, but the other one not,
or both vanish.\footnote{$a_{11} = b_{10} = 0$
but $a_{00} b_{01} \neq 0$.}

If $z_0 = 0$, but $z_1 \neq 0$, then the square root of
(\ref{l2r2.3}) equals $|z_1|^2 \langle 1|\pi|1\rangle$ for
all pure input states. But this can be obviously extended
to a linear function on the input state space. It is easy
to see that there cannot be a larger convex function than
a linear one, if the pre-described values at the pure
states allow its existence. Just that happened with the
degenerate 1-qubit channels. Therefore,
\begin{equation} \label{l2r2.4}
C(\Phi; \omega) = 2 |z_1|^2 \langle 1|\omega|1\rangle \,
\hbox{ if } \, b_{10} a_{00} = 0
\end{equation}
and the Kraus operators are assumed as in (\ref{l2r2.1}).
Similar,
\begin{equation} \label{l2r2.5}
C(\Phi; \omega) = 2 |z_0|^2 \langle 0|\omega|0\rangle \,
\hbox{ if } \, b_{01} a_{11} = 0 .
\end{equation}
Clearly, the concurrence is identical zero if both, $z_0$
and $z_1$, vanish.
\medskip

\noindent Some \underline{$\dim 4 \to \dim 2$ channels}
can be treated which are modifications of the partial
trace. In these cases, $\vartheta$ is proportional to
Wootters' conjugation. Generally, the partial trace
\begin{equation} \label{ptr2}
\T_2 \, X \equiv
\T_2 \, \pmatrix{X_{00} & X_{01} \cr X_{10} & X_{11}}
 =  X_{00} + X_{11} ,
\end{equation}
is of length two and of rank $d$. The construction
(\ref{anti1}) requires $d=2$.

The partial trace can be embedded in a family of
``phase-damping'' channels\footnote{They are trace-preserving
for $q=1/2$ only. See appendix,2},
\begin{equation} \label{ptr5}
\T_{2,q} \, X = X_{00} + X_{11} + (1-2q) (X_{01} + X_{10}) ,
\end{equation}
with $0 < q < 1$ and with Kraus operators
\begin{equation} \label{ptr5a}
A = \sqrt{1-q} \, \pmatrix{\1 & \1}, \quad
B = \sqrt{q} \,  \pmatrix{\1 & - \1}.
\end{equation}
To calculate $\vartheta$ for the channel (\ref{ptr5}), we
start with
\begin{displaymath}
\vartheta =  \sqrt{q(1-q)}  \,
(A_1^* \theta_f A_2 - A_2^* \theta_f A_1) .
\end{displaymath}
We need the Hermitian part of
\begin{displaymath}
\pmatrix{1 & 0 \cr 0 & 1 \cr 1 & 0 \cr 0 & 1}
\pmatrix{0 & -1 \cr 1 & 0}_{\rc}
\pmatrix{1 & 0 & -1 & 0 \cr 0 & 1 & 0 & -1} =
\pmatrix{0 & 1 & 0 & -1 \cr -1 & 0 & 1 & 0 \cr
0 & 1 & 0 & -1 \cr -1 & 0 & 1 & 0 }_{\rc}
\end{displaymath}
An anti-linear operator is Hermitian if every of its
matrix representations is a symmetric matrix. Hence
we obtain, up to a factor, Wootters' conjugation:
\begin{equation} \label{ptr6}
\vartheta =  \sqrt{q(1-q)}  \,
\pmatrix{0 & 0 & 0 & -1 \cr 0 & 0 & 1 &
0 \cr 0 & 1 & 0 & 0 \cr -1 & 0 & 0 & 0 }_{\rc} = \, -
 \sqrt{q(1-q)} \theta_f \otimes \theta_f
\end{equation}
Of course, the same expressions can be deduced by inserting
the matrix entries of (\ref{ptr5a}) in (\ref{matrix2}).
\medskip

Typically one does not know closed expressions for the
concurrence of a channel, but there are estimates,
see \cite{MKB04} for example. An estimation from below
can be obtained for cp-maps of rank two as follows.
Consider the auxiliary maps
\begin{displaymath}
\Phi_{jk}(X) = A_j X A_j^* + A_k X A_k^*,
\end{displaymath}
built with the Kraus operators of $\Phi$.
The following estimate is true:
\begin{equation} \label{estimate}
C(\Phi ; X)^2 \geq \sum_{j < k} C(\Phi_{jk} ; X)^2
\end{equation}
Proof: For $X \geq 0$ of rank one, (\ref{estimate}) becomes
an equality, see (\ref{det2}).
The square root of the right hand side is sub-additive
and homogeneous. By the very definition, the concurrence is
the largest function with these two properties. Hence
(\ref{estimate}) must hold. Similar inequalities, without
the restriction to the rank two case, have been obtained
by Minter et al, \cite{MKB04}.

If $\Phi$ is a cp-map between qubits, then (\ref{estimate})
sharpens to
\begin{equation} \label{estimate1}
C(\Phi ; X)^2 \geq 4 \T \, (X \Phi'(X)) - 8 (\det X)
\sum_{j<k} \sqrt{\det \vartheta_{jk}^2} .
\end{equation}
This can be seen from (\ref{l2r2.3a}), proven later on.

\section{Concurrence}

The aim of the section is to calculate concurrences, a task,
which can be done with satisfaction for length two 1-qubit
channels. In $4 \to 2$ a more explicit discussions seems
possible.

The notion of ``concurrence'' has been explained
already in the introduction. A version, extended to the
positive cone by homogeneity, will be used.
The concept has been developed originally with
respect to partial traces \cite{Woo01}.
However, by the Stinespring dilatation theorem any
trace-preserving cp-map is equivalent to a sub-channel
of a partial trace.
\medskip

\underline{Definition}
Let $\Phi$ be a positive map of rank two. $C(\Phi ; X)$, the
``$\Phi$-concurrence'', is defined for all positive
operators $X$ of the input space by the following
properties:\\
(i) \, $C(\Phi ; X)$ is homogeneous of degree one,
\begin{displaymath}
C(\Phi ; \lambda X) = \lambda \, C(\Phi ; X)  ,
\quad \lambda \geq 0 .
\end{displaymath}
(ii) \, $C(\Phi ; X)$ is sub-additive,
\begin{displaymath}
C(\Phi ; X + Y) \leq C(\Phi ; X) + C(\Phi ; Y)
\end{displaymath}
(iii) \, $C(\Phi ; X)$ is the largest function with properties
(i) and (ii) above, satisfying for all vectors $\psi$ of the
input space
\begin{equation} \label{conc1}
C(\Phi ; | \psi \rangle\langle \psi |) = 2
\sqrt{\det \Phi(| \psi \rangle\langle \psi | )}
\end{equation}

There are other, equivalent possibilities to define $C$.
One knows
\begin{equation} \label{conc2}
C(\Phi ; X) = 2 \inf
\{ \sum \sqrt{\det \Phi(|\psi_j \rangle\langle \psi_j|) }, \quad
\sum |\psi_j \rangle\langle \psi_j| = X \} .
\end{equation}
Next, just because the square root of the determinant
is concave in dimension two, the convex hull construction
applies,
\begin{equation} \label{conc3}
C(\Phi ; X) = 2 \inf
\{ \sum \sqrt{\det \Phi(X_j) }, \quad
\sum X_j = X \} ,
\end{equation}
so that the $X_j \geq \0$ can be arbitrarily chosen up to the
constraint of summing up to $X$. Notice, that a similar trick
with the determinant (or the second symmetric function) in
the definition of concurrence would fail because the
determinant is neither concave nor homogeneous on the cone
of positive operators.
\medskip

For cp-maps of rank and length two much can be said about
the variational problem involved in the definitions above.
This is due to the fact that the derived Kraus space is
1-dimensional, and there is only one $\vartheta$ as
explained in the previous section.
The appropriate extension of the procedure invented by
Wootters is in \cite{Uh00c} and it goes this way:\\
{\em Step 1.}
For two positive operators, $X_1$ and $X_2$,
of the input space we need
\begin{equation} \label{conc4}
\{ \lambda_1 \geq \lambda_2 \geq \dots \} =
\hbox{ eigenvalues of } (X_1^{1/2} X_2 X_1^{1/2})^{1/2}
\end{equation}
to define
\begin{equation} \label{conc5}
C(X_1 , X_2) := \max \{ 0, \, \lambda_1 - \sum_{j > 1} \lambda_j \} .
\end{equation}
{\em Step 2.} We set $X = X_1$ and replace
$X_2$ by $\vartheta X \vartheta$,
\begin{equation} \label{conc6}
C(\Phi ; X) = 2 \, C(X , \vartheta X \vartheta) ,
\end{equation}
and we are done. The proof is in \cite{Uh00c}.

It follows from its definition that the restriction of
$C(\Phi;.)$ onto the intersection of the cone of positive
$X$ with an affine hyperplane, $\T (X_0X)=1$, with a given
invertible $X_0$, is a convex roof. It is the largest convex
function attaining values given on the rank one operators
contained in the intersection.
\medskip

\noindent \underline{Example}. In treating the modified partial
trace $\T_{2,q}$ of (\ref{ptr5}), we had computed in (\ref{ptr6})
$\vartheta = -\sqrt{q(1-q)}\theta_w$. Here
$\theta_w = -\theta_f \otimes \theta_f$ is Wootters
conjugation. We conclude by homogeneity
\begin{displaymath}
C(\T_{2,q} ; X) = 4 \sqrt{q(1-q)} C(\T_2 ; X)
\end{displaymath}
and the right hand side is Wootters' concurrence
in \cite{Woo97}. Therefore, the optimal decompositions
of the modified partial traces (\ref{ptr5}) do not depend
on $q$, $0 < q < 1$.

\noindent \underline{Remark}.  Fei et al \cite{FJLW03}
have pointed out
a class of states allowing for calculating concurrence
by arriving at an analogue of (\ref{conc5}).
Their ``computable'' density
operators come with two different eigenvalues of equal
degeneracy. The authors use $(\det Y)^{1/d}$, which is
concave, and  which becomes a quadratic form for their
states. $Y$ stands for the partial trace of the input.

\section{1-qubit channels of length two.}
Due to the presence of only two
eigenvalues, $\lambda_1, \lambda_2$, in (\ref{conc4})
one can get a more detailed picture:
The right hand side of
(\ref{conc5}) becomes $\lambda_1 - \lambda_2$. Combining
\begin{displaymath}
(\lambda_1 - \lambda_2)^2 = (\T \, \xi)^2 - 4 \det \xi,
\quad \xi = (X_1^{1/2} X_2 X_1^{1/2})^{1/2}
\end{displaymath}
with the characteristic equation
\begin{displaymath}
(\T \, \xi)^2 = \T \, \xi^2  + 2 \det \xi ,
\end{displaymath}
yields
\begin{displaymath}
(\lambda_1 - \lambda_2)^2 =  \T \, \xi^2 - 2 \det \xi .
\end{displaymath}
Finally, removing the auxiliary operator $\xi$, we obtain
\begin{equation} \label{conc5a}
C(X_1, X_2)^2 = \T \, (X_1 X_2) - 2 \sqrt{\det (X_1 X_2)}  .
\end{equation}

Let $\Phi$ be the cp-map with the Kraus operators $A, B$ of
(\ref{l2r2.1}). We have to substitute $X = X_1$ and
$X_2 = \vartheta X \vartheta$ into (\ref{conc5a}),
remembering (\ref{conc6}):
\begin{equation} \label{l2r2.3a}
{1 \over 4} \, C(\Phi ; X)^2 = \T \, (X \vartheta X \vartheta)
- 2 (\det X) \, (\det \vartheta^2)^{1/2} .
\end{equation}
$\vartheta$ is taken from (\ref{l2r2.2}). It is
diagonal in the reference basis with entries $z_0^2$ and
$-z_1^2$. We arrive at ($\to$ see appendix,3)
\begin{displaymath}
\T \, X \vartheta X \vartheta =
  (z_0^* x_{00} z_0)^2 - (z_0^* x_{01} z_1)^2
- (z_0 x_{10} z_1^*)^2 + (z_1^* x_{11} z_1)^2 ,
\end{displaymath}
\begin{displaymath}
(\det X) \, (\det \vartheta^2)^{1/2} =
(z_0 z_0^* z_1 z_1^*)(x_{00} x_{11} - x_{01} x_{10}) .
\end{displaymath}
Combining these two expressions as dictated by
(\ref{l2r2.3a}) results in
\begin{equation} \label{l2r2.6}
C(\Phi ; X)^2 = 4 (z_0 z_0^* x_{00} - z_1 z_1^* x_{11})^2
- 4 (z_0 z_1^* x_{10} - z_1 z_0^* x_{01})^2 .
\end{equation}
The number within the second delimiter is purely
imaginary and, therefore, {\em $C^2$ is the sum of
two positive quadratic terms.}
This observation remains true if we allow for any
Hermitian operator in (\ref{l2r2.6}).

{\bf Lemma 2.} \,
The squared concurrence (\ref{l2r2.6}) is a
positive semi-definite quadratic form
on the real-linear space of Hermitian Operators.\\
The concurrence is a Hilbert semi-norm.$\diamond$
\medskip

There is a further remarkable observation:
The concurrence (\ref{l2r2.6})
is equal to the absolute value of the complex number
\begin{equation} \label{l2r2.7}
c(X) := 2 (z_0 z_0^* x_{00} - z_1 z_1^* x_{11} +
z_0 z_1^* x_{10} - z_1 z_0^* x_{01}) .
\end{equation}
The imaginary part vanishes if and only if $\Phi$
becomes degenerate. Let now us rewrite (\ref{l2r2.7})
for Hermitian $X$ as follows
\begin{equation} \label{line1}
C^2(\Phi; X)^2 = l_1^2(X) + l_2^2(X)
\end{equation}
by the help of the real linear forms
\begin{equation} \label{line2}
l_1(X) = 2(z_0 z_0^* x_{00} - z_1 z_1^* x_{11}), \quad
l_2(X) = 2i (z_0 z_1^* x_{10} - z_1 z_0^* x_{01}) .
\end{equation}
$l_2$ remains constant along
\begin{equation} \label{line3}
x_{01}' = z_0 z_1^* t + x_{01}, \quad
x_{10}' = z_0^* z_1 t + x_{10}
\end{equation}
and only the off-diagonal entries of the input operator
vary. The values of $C$ and of $l_1$ determine $l_2$
and, hence, the diagonal elements of the input operator.
Therefore, we may rewrite (\ref{line3}) to
\begin{displaymath}
 X' = X + t \, \pmatrix{0 & z_0 z_1^* \cr z_0^* z_1 & 0} .
\end{displaymath}
We can relax from the condition that the traces of
$X$ and $X'$ are equal. Indeed, the concurrence
remain constant on the planes
\begin{displaymath}
 X' = X + t \, \pmatrix{0 & z_0 z_1^* \cr z_0^* z_1 & 0} .
+ \hat t \, \pmatrix{z_1 z_1^* & 0 \cr 0 & z_0 z_0^*} ,
\end{displaymath}
or, equivalently,
\begin{displaymath}
X' = X
+ t_1 \pmatrix{z_1 z_1^* & z_0 z_1^* \cr z_1 z_0^* & z_0 z_0^*}
+ t_2 \pmatrix{z_1 z_1^* & - z_0 z_1^* \cr - z_1 z_0^* & z_0 z_0^*}.
\end{displaymath}
The two vectors
\begin{equation} \label{line4}
\psi_1 = z_1^* |0\rangle + z_0^* |1\rangle, \quad
\psi_2 = z_1^* |0\rangle - z_0^* |1\rangle ,
\end{equation}
are solutions of $\langle \psi, \vartheta \psi \rangle = 0$,
and represent two linear independent $\Phi$-separable
vectors. ($\to$ See also appendix,4.)

{\bf Lemma 3} \,
The concurrence of a 1-quibt cp-map $\Phi$ with $\Phi$-separable
vectors $\psi_1$ and $\psi_2$ is constant on every plane
\begin{equation} \label{line5}
X' = X + t_1 |\psi_1 \rangle\langle \psi_1| +
t_2 |\psi_2 \rangle\langle \psi_2|
\end{equation}
with $X$ Hermitian and $t_1$, $t_2$ real.$\diamond$
\medskip

We have seen that every mixed state is on a straight
line of constant concurrence, and that line is unique
in the non-degenerate case. It then hits the Bloch sphere
at exactly two pure states. Let us look at this family of
parallel lines in Bloch space. It is geometrically evident
that their must be a reflection on a plane perpendicular
to these lines which reflects the Bloch ball onto
itself. Such a reflection cannot be unitary, because it
changes the orientation of the Bloch ball. That is, we
ask for a conjugation implementing the said reflection.

For the computation we assume $\Phi$ non-degenerate.
Given an Hermitian $X$, we look for a change leaving
the number $c(X)$ of (\ref{l2r2.7}) and the linear forms
(\ref{line2}) invariant. This is achieved by
\begin{displaymath}
x_{01} \to - \frac{z_0 z_1^*}{z_0^* z_1} x_{10}, \quad
x_{10} \to - \frac{z_1 z_0^*}{z_1^* z_0} x_{01}
\end{displaymath}
and by letting the diagonal of $X$ unchanged. Then Trace
and Determinant of $X$ are invariant and the Bloch sphere
is mapped onto itself. This suggests that
\begin{equation} \label{line6}
\theta (c_0 |0\rangle + c_1 |1\rangle) =
c_0^* \frac{z_0}{z_0^*} |0\rangle -
c_1^* \frac{z_1}{z_1^*} |1\rangle
\end{equation}
is the conjugation we are looking for.  Indeed, starting
with any matrix $X$, one arrives after a
straightforward calculation  at
\begin{equation} \label{line7}
\theta X^* \theta =
\pmatrix{x_{00} & \epsilon x_{10} \cr
\epsilon^* x_{01} & x_{11}}, \quad
\epsilon = - (z_0 z_1^*)(z_0^* z_1)^{-1} .
\end{equation}
Therefore, (\ref{line6}) is the desired conjugation which
transforms the Bloch space onto itself and
does not change $c(X)$. This proves the main part of

{\bf Theorem 2} \,
Let $\Phi$ be a non-degenerate 1-qubit map of length two.
Define $\theta$ by the polar decomposition
\begin{equation} \label{line8}
\vartheta = \theta \, |\vartheta| = |\vartheta| \, \theta,
\quad |\vartheta| = (\vartheta^2)^{1/2}, \quad
\vartheta = \pmatrix{z_0/z_0^* & 0 \cr 0 & - z_1/z_1^*}_{\rc} \; .
\end{equation}
$\theta$ is a conjugation satisfying
\begin{equation} \label{line9}
c(\theta X^* \theta) = c(X) .
\end{equation}
The transformation $X \to \theta X^* \theta$ maps every
line of constant concurrence into itself.$\diamond$
\medskip

It remains to establish (\ref{line8}). Because the operators
are diagonal in the reference basis, the assertion reduces to
\begin{displaymath}
z_0^2 = \frac{z_0}{z_0^*} \, |z_0^2|, \quad
z_1^2 = - \frac{z_1}{z_1^*} \, |z_1^2|,
\end{displaymath}
which is obviously true.

Next we construct a further conjugation, $\theta'$, operating
on the out-operators. It would be appropriate, to call the
previous constructed one $\theta^{\rm in}$ and the one yet
to be defined $\theta^{\rm out}$. However, we use simply
$\theta$ and $\theta'$, not to overload our equations.
The geometric meaning of $\theta'$ is similar to that of
$\theta$. $\Phi$ maps the parallel lines of constant
concurrence onto a family of parallel lines of the output
states. $\theta'$ transforms every such output line
into itself. As it
must interchange the outputs of the $\Phi$-separable states,
the line through these two pure states determines the
output family of lines completely.  Hence, $\theta'$ is fixed
up to a phase factor.

To begin with, we remember (\ref{l2r2.1}) and
introduce the uni-modular numbers
\begin{equation} \label{line10}
\epsilon_0' = - \frac{a_{00} b_{01}}{|a_{00} b_{01}|},
\quad
\epsilon_1' =  \frac{a_{11} b_{10}}{|a_{11} b_{10}|} .
\end{equation}
We are in the position to introduce $\theta'$.
\begin{equation} \label{line11}
\theta' ( c_0 |0\rangle + c_1 |1\rangle ) =
\epsilon_0' c_0^* |0\rangle + \epsilon_1' c_1^* |1\rangle
\end{equation}
A rather straightforward calculation yields
($\to$ See appendix, 5.)

{\bf Lemma 4.} \,
Let $A$, $B$, be the Kraus operators (\ref{l2r2.1}) of $\Phi$.
Then
\begin{equation} \label{line12}
\theta' A \theta = \frac{b_{01} b_{10}}{|b_{01} b_{10}|} A,
\quad
\theta' B \theta = - \frac{a_{00} a_{11}}{|a_{00} a_{11}|} B ,
\end{equation}
and, therefore
\begin{equation} \label{line13}
\theta' \, \Phi( \theta X \theta) \, \theta' = \Phi(X) .
\end{equation}
\medskip

\section{Entanglement with respect to $\Phi$}

Again, the essence of what is following goes back to
\cite{BenFucSmo96} and \cite{Woo97}, see also
\cite{Uh98a}, appendix, \cite{BNU02}, and
\cite{Uh03c} for a short introduction to roofs.

The definition of $E(\Phi;.)$, mentioned in the introduction,
can be extended to the positive cone. At first we extend the
entropy of output states by scaling. The ``scaled von Neumann
entropy'' reads
\begin{equation} \label{sentropy}
S_{\rm sc}(Y) = [S(\T Y)] \, S(Y / [S(\T Y)]) =
\eta(Y) - \eta(\T Y)
\end{equation}
with $\eta(y) = -y \log y$. On the state space,
$S_{\rm sc}$ is the usual von Neumann entropy.
For positive $Y$ (\ref{sentropy})
provides super-additivity and homogeneity,
\begin{equation} \label{sentropy1}
S_{\rm sc}(Y_1 + Y_2) \geq S_{\rm sc}(Y_1) + S_{\rm sc}(Y_2),
\quad \lambda S_{\rm sc}(Y) = S_{\rm sc}(\lambda Y) .
\end{equation}
Now we can proceed similar as in Definition 3.
\medskip

\underline{Definition} \,
Let $\Phi$ be a positive map of rank two. $E(\Phi ; X)$, the
``$\Phi$-entanglement'', is the largest function on the
positive cone of the input system fulfilling
\begin{displaymath}
E(\Phi ; X_1 + X_2) \leq E(\Phi ; X_1) + E(\Phi ; X_2) ,
\end{displaymath}
\begin{displaymath}
\lambda E(\Phi ; X) = E(\Phi ; \lambda X), \quad \lambda \geq 0,
\end{displaymath}
\begin{equation} \label{etf1}
{\rm rank}(X) = 1  \longrightarrow \,
E(\Phi ; X) = S_{\rm sc}(\Phi(X)) .
\end{equation}
\medskip

The definition reduces to the
one addressed in the introduction for channels. Alternatively
one may use all decompositions of $X$ with positive summands,
\begin{equation} \label{etf2}
E(\Phi ; X) = \inf \sum S_{\rm sc}(X_j), \quad X = \sum X_j .
\end{equation}
Let us now return to our particular case of a cp-map of rank
two and of length two. Then
\begin{displaymath}
\T Y = 1 \rightarrow S_{\rm sc}(Y) =
\eta([1 + \sqrt{1 - 4 \det Y}]/2) + \eta([1 - \sqrt{1 - 4 \det Y}]/2) .
\end{displaymath}
With $Y = \T \Phi(X)$ and rank$(X)=1$ this coincides with
\begin{equation} \label{etf3}
\eta([1 + \sqrt{1 -  C(\Phi; X)^2 }]/2) +
\eta([1 - \sqrt{1 - C(\Phi; X)^2 }]/2) .
\end{equation}
One knows already from \cite{BenFucSmo96}, \cite{Woo97},
\cite{Uh00c}, this a convex function. Assuming
\begin{displaymath}
\T \, \Phi(X) = \T \,X_0 X, \quad \det X_0 \neq 0 ,
\end{displaymath}
the restriction of $C(\Phi;X)$ to $\T \Phi(X) = 1$ becomes
a convex roof. Being flat, every optimal decomposition of
$C$ remains optimal for (\ref{etf3}). Therefore, it coincides
with $E(\Phi;X)$ if restricted to $\T \Phi(X)=1$. However,
by homogeneity, it must be true for all $X \geq 0$.
That is the content of

{\bf Theorem 3.} \,
Let $\Phi$ be completely positive, trace preserving, of rank
two, and with Kraus operators $A$ and $B$. Assume
$A^*A + B^*B$ invertible. Then
\begin{displaymath}
E(\Phi ; X) = \eta(y_+) + \eta(y_-) - \eta(y_+ + y_-)
\end{displaymath}
\begin{equation} \label{etf4}
2 y_{\pm} = \T \Phi(X) \pm \sqrt{ [\T \Phi(X)]^2 - C(\Phi ; X)^2} .
\end{equation}
\medskip

The theorem allows for a fairly explicit expression for
maximized Holevo quantities. For a channel $\Phi$ and an
ensemble of states of the input space, Holevo's quantity is
\begin{displaymath}
\chi = S(\Phi(\omega)) - \sum p_j S(\Phi(\omega_j))
\end{displaymath}
with $\omega$ the average of the $\omega_j$ with weights $p_j$.
Being states, nothing changes in replacing $S$ by the scaled
von Neumann entropy. But because of the homogeneity, we can
write
\begin{displaymath}
\chi = S_{\rm sc}(\Phi(\omega)) -
\sum S_{\rm sc}(\Phi(p_j \omega_j))
\end{displaymath}
Given $\omega$, the ``maximized Holevo quantity'' is the supreme
$\chi^*$ of $\chi$ if one runs through all ensembles with
average $\omega$. By homogeneity we need not respect normalization.
Thus
\begin{equation} \label{holevo1}
\chi^*(\Phi ; X) = S_{\rm sc}(\Phi(X)) - E(\Phi ; X), \quad
X \geq 0 .
\end{equation}
is a concave function on the positive input operators, identical
with the usual $\chi^*$ for density operators and channels
$\Phi$.

We now return to the 1-qubit channel. We already have
computed $E$, so that we have (\ref{holevo1}) as a function
of $X$, built from logarithms and algebraic terms.
We can do even better. For non-degenerate $\Phi$
we can rely on lemma 7 to see that
 both terms in (\ref{holevo1}) are
$\theta$-invariant, and not only $E$. For positive $X$
we obtain
\begin{equation} \label{holevo2}
\chi^*(\Phi ; X) = \chi^*(\Phi ; \theta X \theta) .
\end{equation}
To get the Holevo capacity, we have to maximize (\ref{holevo2})
over all density operators. $\theta$ is a symmetry of this
set. The concavity of (\ref{holevo2}) guaranties that there
must be a $\theta$-invariant state at which the maximum is
attained. Therefore, it suffices to search in the set
of all $\omega = \theta \omega \theta$ . (\ref{line7})
provides the conditions for $\theta$-invariance.

{\bf Lemma 5.} \, The maximum
\begin{equation} \label{holevo3}
\chi^*(\Phi) = \max_{omega} \chi(\Phi ; \omega), \quad
\omega \hbox{ density operator}
\end{equation}
is attained on a $\theta-invariant$ state.
Assuming (\ref{l2r2.1}) and denoting by
$\omega_{jk}$ the matrix entries of $\omega$,
then $\omega$ belongs to plane given by
\begin{displaymath}
z_0^* \omega_{01} z_1 + z_1^* \omega_{10} z_0 = 0,
\end{displaymath}
i.~e. if $z_0^*\omega_{01}z_1$ is purely imaginary.$\diamond$

See also appendix, 6.
\medskip

In the degenerate case, the
search for the maximum (\ref{holevo3}), i.~e.~for
the Holevo capacity, can even be done on a line in
Bloch space, see \cite{Uh01}: The concurrence,
(\ref{l2r2.4}) or (\ref{l2r2.5}),
becomes constant on planes, and there is a line,
perpendicular to the planes, on which the maximum
is to search.

\section{Appendix}
The appendix is added to provide further explanations and
calculations to the main text.
\medskip

{\bf 1.} \,
In any Hilbert space one defines the Hermitian adjoint
$\vartheta^*$ of an anti-linear operator $\vartheta$ by
\begin{displaymath}
\langle \psi_1, \vartheta^* \psi_2 \rangle =
\langle \psi_2, \vartheta \psi_1 \rangle .
\end{displaymath}
In particular, $\theta_f^* = \theta_f^{-1} = - \theta_f$.

Let us look more detailed at (\ref{matrix2}) assuming that
we are in the 2x2-case. We find
\begin{displaymath}
A^* \theta_f B = A^* \pmatrix{0 & 1 \cr -1 & 0}_{\rc}
\pmatrix{b_{00} & b_{01} \cr b_{10} & b_{1}} = A^*
\pmatrix{b_{10}^* & b_{11}^* \cr -b_{00}^* & -b_{01}^*}_{\rc} ,
\end{displaymath}
\begin{displaymath}
A^* \theta_f B =
\pmatrix{a_{00^*} & a_{10}^* \cr a_{01}^* & a_{11}^*}
\pmatrix{b_{10}^* & b_{11}^* \cr -b_{00}^* & -b_{01}^*}_{\rc}
\end{displaymath}
and, finally,
\begin{displaymath}
A^* \theta_f B = \pmatrix{
a_{00}^* b_{10}^* - a_{10}^* b_{00}^* &
a_{00}^* b_{11}^* - a_{10}^* b_{01}^* \cr
a_{01}^* b_{10}^* - a_{11}^* b_{00}^* &
a_{01}^* b_{11}^* - a_{11}^* b_{01}^* }_{\rc} \; .
\end{displaymath}
To get the Hermitian part, we have to remember
$\theta_f^* = - \theta_f$. An anti-linear operator is Hermitian
iff its matrix representation in every basis is a symmetric
matrix. Hence we get (\ref{matrix2}) in the 2x2-case.
Completely similar one get (\ref{matrix2}) in general.
\medskip

{\bf 2.a.} \, We presently consider determinants, not the
trace or the action onto the unit operator $\1$.

To have trace preserving with Kraus operators (\ref{l2r2.1}),
one has to require
\begin{equation} \label{ptr}
a_{00}^* a_{00} + b_{10}^* b_{10} = 1 , \quad
a_{11}^* a_{11} + b_{01}^* b_{01} = 1
\end{equation}
because $A^* A + B^* B = \1$ is satisfied then.

If the map becomes unital if $AA^* + BB^* = \1$, hence
\begin{equation} \label{pun}
a_{00}^* a_{00} + b_{01}^* b_{01} = 1 , \quad
a_{11}^* a_{11} + b_{10}^* b_{10} = 1
\end{equation}
A doubly stochastic channel satisfies both, (\ref{ptr})
and (\ref{pun}). Hence we get
\begin{equation} \label{ptrun}
|a_{00}| = |a_{11}| = |b_{01}| = |b_{10}|
\end{equation}
necessarily. (\ref{ptrun}) becomes sufficient for
bi-stochasticity if one of the four equations in
(\ref{ptr}) and (\ref{pun}) is valid.
\medskip

{\bf 2.b.} \, Within the maps $\T_{2,q}$, defined in
(\ref{ptr5}), only the partial trace the is trace preserving.
However, by restricting to the linear space of block-matrices
with $\T (X_{01} + X_{10}) = 0)$, we get a channel.

On the other hand, the maps $(1/2) \T_{2,q}$
with Kraus operators (\ref{ptr5a}) are unital.
Hence their duals
\begin{displaymath}
X \to A^* X A + B^* X B
\end{displaymath}
are trace preserving maps from one qubit into two qubit states.

Generally, all maps with Kraus operators taken from the
Kraus space spanned by $A^*$, $B^*$ as given in (\ref{ptr5a})
can be described as following:\\
There is a positive 2x2-matrix with entries $\alpha_{jk}$
such that
\begin{displaymath}
X \mapsto \pmatrix{ \alpha_{00} X & \alpha_{01} X \cr
\alpha_{10} X & \alpha_{11} X } \; .
\end{displaymath}
\medskip

{\bf 3.} \,
Let us do the calculation in more detail. While (\ref{l2r2.6})
assumes $X$ Hermitian, we take a general $X$ here and consider
at first $\T \, X \vartheta X^* \vartheta$ and start with
\begin{displaymath}
X^* \vartheta =
\pmatrix{x_{00}^* & x_{10}^* \cr x_{01}^* & x_{00}^*}
\pmatrix{z_0^2 & 0 \cr 0 & -z_1^2}_{\rc} =
\pmatrix{z_0^2 x_{00}^* & -z_1^2 x_{10}^* \cr
z_0^2 x_{01}^* & -z_1^2 x_{00}^*}_{\rc} ,
\end{displaymath}
\begin{displaymath}
\vartheta X^* \vartheta = \pmatrix{z_0^2 & 0 \cr 0 & -z_1^2}_{\rc}
\pmatrix{z_0^2 x_{00}^* & -z_1^2 x_{10}^* \cr
z_0^2 x_{01}^* & -z_1^2 x_{00}^*}_{\rc} =
\pmatrix{(z_0 z_0^*)^2 x_{00} & -(z_0 z_1^*)^2 x_{10} \cr
-(z_1 z_0^*)^2 x_{01} & (z_1 z_1^*)^2 x_{11} }
\end{displaymath}
resulting in
\begin{displaymath}
\T \, X \vartheta X^* \vartheta = \T \,
\pmatrix{x_{00} & x_{01} \cr x_{10} & x_{11}}
\pmatrix{(z_0 z_0^*)^2 x_{00} & -(z_0 z_1^*)^2 x_{10} \cr
-(z_1 z_0^*)^2 x_{01} & (z_1 z_1^*)^2 x_{11} }
\end{displaymath}
and, as asserted in the main text for Hermitian $X$,
\begin{displaymath}
\T \, X \vartheta X^* \vartheta =
  (z_0^* x_{00} z_0)^2 - (z_0^* x_{01} z_1)^2
- (z_0 x_{10} z_1^*)^2 + (z_1^* x_{11} z_1)^2 ,
\end{displaymath}

We now calculate
\begin{displaymath}
\det X \vartheta X^* \vartheta =
(\det X) \det (\vartheta X^* \vartheta)
\end{displaymath}
and further
\begin{displaymath}
\det (\vartheta X^* \vartheta) = |z_0 z_1|^4 \det X \; ,
\end{displaymath}
\begin{displaymath}
\det X \vartheta X^* \vartheta = |z_0 z_1|^4
(x_{00} x_{11} - x_{01} x_{10})^2 \; .
\end{displaymath}

Now we consider $(1/4)C^2$ as given by (\ref{l2r2.3a}).
One gets
\begin{displaymath}
|z_0 z_1|^2 (\frac{z_0 z_0^*}{z_1 z_1^*} x_{00}^2  +
\frac{z_1 z_1^*}{z_0 z_0^*} x_{11}^2 -
\frac{z_1 z_0^*}{z_0 z_1^*} x_{01}^2 -
\frac{z_0 z_1^*}{z_1 z_0^*} x_{10}^2 -
2 x_{00} x_{11} - 2 x_{01} x_{10})
\end{displaymath}
and it can be rewritten into the form
\begin{displaymath}
|z_0 z_1|^2 \bigl(
(|\frac{z_0}{z_1}| x_{00} - |\frac{z_1}{z_0}| x_{11} )^2 -
(\sqrt{\frac{z_1 z_0^*}{z_0 z_1^*}} x_{01} -
\sqrt{\frac{z_0 z_1^*}{z_1 z_0^*}} x_{10})^2 \bigr)
\end{displaymath}
or, equivalently,
\begin{displaymath}
(z_0 z_0^* x_{00} - z_1 z_1^* x_{11})^2 - z_0 z_0^* z_1 z_1^*
(\sqrt{\frac{z_1 z_0^*}{z_0 z_1^*}} x_{01} -
\sqrt{\frac{z_0 z_1^*}{z_1 z_0^*}} x_{10})^2  \, .
\end{displaymath}
Sign changes in $z_0$ and $z_1$ do not affect
the expression above, moreover

\begin{equation} \label{signconv1}
\sqrt{\frac{z_1 z_0^*}{z_0 z_1^*}} = \frac{z_1 z_0^*}{|z_1 z_0|}
, \quad \sqrt{\frac{z_0 z_1^*}{z_1 z_0^*}}
= \frac{z_0 z_1^*}{|z_1 z_0|} \; .
\end{equation}
In terms of the matrix elements of the Kraus operators
(\ref{l2r2.1}) we have
\begin{equation} \label{signconv2}
\frac{z_1 z_0^*}{|z_1 z_0^*|} = \sqrt{
\frac{a_{00} b_{01}^* b_{10} a_{11}^*}{|a_{00} b_{01} b_{10} a_{11}|}
} , \quad \frac{z_0 z_1^*}{|z_1 z_0^*|} = \sqrt{
\frac{a_{00}^* b_{01} b_{10}^* a_{11}}{|a_{00} b_{01} b_{10} a_{11}|}
} \; .
\end{equation}
\medskip

{\bf 4.} \, According to (\ref{det2}) it is
\begin{displaymath}
\det \Phi(|\psi \rangle\langle \psi|) =
| \langle \psi , \vartheta \psi \rangle |^2
\end{displaymath}
and $\langle \psi, \vartheta \psi \rangle = 0$ is the equation
for the two $\Phi$-separable vectors. Up to normalization these
vectors can be written as in (\ref{line4}). Now
(\ref{line6}) is equivalent with
\begin{displaymath}
\theta \pmatrix{ c_0 \cr c_1 } = \pmatrix{ (z_0 / z_0^*)
c_0^* \cr - (z_1 / z_1^*) c_1^* }, \quad
\theta = \pmatrix{z_0/z_0^* & 0 \cr 0 &
- z_1/z_1^*}_{\rc}
\end{displaymath}
A particular case reads
\begin{displaymath}
\theta |\psi_1\rangle = \frac{z_0 z_1}{z_0^* z_1^*} |\psi_2\rangle,
\quad
\theta |\psi_2\rangle = \frac{z_0 z_1}{z_0^* z_1^*} |\psi_1\rangle \; .
\end{displaymath}
Let us also mention
\begin{displaymath}
\theta \pmatrix{0 & z_0 z_1^* \cr z_0^* z_1 & 0} \theta =
- \pmatrix{0 & z_0 z_1^* \cr z_0^* z_1 & 0} \; .
\end{displaymath}
Similar as in part 3 above we get
\begin{equation} \label{hilf}
\pmatrix{\delta_0 & 0 \cr 0 & \delta_1}_{\rc} X^*
\pmatrix{\delta_0 & 0 \cr 0 & \delta_1}_{\rc} =
\pmatrix{(\delta_0 \delta_0^* x_{00} & \delta_0 \delta_1^* x_{10} \cr
\delta_1 \delta_0^* x_{01} & \delta_1 \delta_1^* x_{11} }
\end{equation}
and we specify this equation to
\begin{displaymath}
\theta X^* \theta
=
\pmatrix{ x_{00} & -((z_0 z_1^*)/(z_0^* z_1)) x_{10} \cr
-  ((z_0^* z_1)/(z_0 z_1^*)) x_{01} & x_{11} } \; .
\end{displaymath}
It follows
\begin{displaymath}
X - \theta X^* \theta =
\pmatrix{0 & x_{01} + ((z_0 z_1^*)/(z_0^* z_1)) x_{10} \cr
 x_{10} + ((z_0^* z_1)/(z_0 z_1^*)) x_{01} & 0}
\end{displaymath}
and
\begin{equation} \label{hilf1}
X - \theta X^* \theta =
\bigl( \frac{x_{01}}{z_0 z_1^*} + \frac{x_{10}}{z_0^* z_1}
\bigr) \pmatrix{0 & z_0 z_1^* \cr z_0^* z_1 & 0} \; .
\end{equation}
If $X$ is Hermitian, the factor at the right hand side
is real.\\
{\em If $X = X^*$, then the two lines}
\begin{displaymath}
X + \pmatrix{0 & z_0 z_1^* \cr z_0^* z_1 & 0} \bbbr
\, \hbox{ and } \, \theta X \theta +
\pmatrix{0 & z_0 z_1^* \cr z_0^* z_1 & 0} \bbbr
\end{displaymath}
{\em are equal.} This proves the last assertion in
theorem 2. Similar it is with Lemma 3: {\em If a plane
(\ref{line5}) consists of Hermitian matrices then
it is transformed onto itself by $X \to \theta X \theta$.}

Let us also note: If a line (plane) of constant concurrence
contains a density operator, it is transformed onto itself
by $X \to \theta X \theta$.
\medskip

{\bf 5.} \, We may write
\begin{displaymath}
\theta' = \pmatrix{\epsilon_0' & 0 \cr 0 & - \epsilon_1'}_{\rc}
\end{displaymath}
in a basis where (\ref{l2r2.1}) is valid. Here,
see (\ref{line10}),
\begin{displaymath}
\epsilon_0' =  \frac{a_{00} b_{01}}{|a_{00} b_{01}|},
\quad
\epsilon_1' =  \frac{a_{11} b_{10}}{|a_{11} b_{10}|} .
\end{displaymath}
We also have
\begin{displaymath}
\theta = \pmatrix{\epsilon_0 & 0 \cr 0 & - \epsilon_1}_{\rc}
\end{displaymath}
with
\begin{displaymath}
\epsilon_0 = (z_0 / z_0^*) =
\frac{b_{10}^* a_{00}^*}{|b_{10} a_{00}|} .
\quad
\epsilon_1 = (z_1 / z_1^*) =
\frac{b_{01}^* a_{11}^*}{|b_{01} a_{11}|} .
\end{displaymath}
Let us compute
\begin{displaymath}
A \theta = \theta' \pmatrix{a_{00} & 0 \cr 0 & a_{11}}
\pmatrix{\epsilon_0 & 0 \cr 0 & - \epsilon_1}_{\rc} =
\pmatrix{\epsilon_0 a_{00} & 0 \cr 0 & -\epsilon_1 a_{11}}_{\rc}
\end{displaymath}
\begin{displaymath}
\theta' A \theta =
\pmatrix{\epsilon_0' & 0 \cr 0 & - \epsilon_1'}_{\rc}
\pmatrix{\epsilon_0 a_{00} & 0 \cr 0 & -\epsilon_1 a_{11}}_{\rc}
= \pmatrix{\epsilon_0' a_{00}^* \epsilon_0^* & 0 \cr 0 &
\epsilon_1' a_{11}^* \epsilon_1^* }
\end{displaymath}
Therefore, the non zero entries of $\theta' A \theta$ are
\begin{displaymath}
\frac{a_{00}^2 a_{00}^*b_{01} b_{10}}{|a_{00}^2 b_{01} b_{10}|}
\, \hbox{ and } \,
\frac{a_{11}^2 a_{11}^*b_{01} b_{10}}{|a_{11}^2 b_{01} b_{10}|}
\end{displaymath}
which proves
\begin{displaymath}
\theta' A \theta = \frac{b_{01} b_{10}}{|b_{01} b_{10}|} A \; .
\end{displaymath}
Similarly one gets the other relation in (\ref{line12}): One
can first compute
\begin{displaymath}
\theta' B \theta = - \pmatrix{0 & \epsilon_0 b_{01}^* \epsilon_1^*
\cr \epsilon_1 b_{10}^* \epsilon_0^*  }
\end{displaymath}
and then show that non vanishing the entries are
\begin{displaymath}
\frac{a_{00} a_{11} b_{01}^2 b_{01}^*}{|a_{00} b_{01}^2 a_{11}|}
\, \hbox{ and } \,
\frac{a_{00} a_{11} b_{10}^2 b_{10}^*}{|a_{00}^2 b_{10}^2 a_{11}|}
\; .
\end{displaymath}
This way (\ref{line12}) has been proved. But (\ref{line12})
implies (\ref{line13}).

Let us define in analogy to (\ref{line4})
\begin{equation} \label{line14}
\psi_1' = \sqrt{a_{00} b_{01}} |0\rangle +
\sqrt{a_{11} b_{10}} |1\rangle,
\quad
\psi_2' = \sqrt{a_{00} b_{01}} |0\rangle -
\sqrt{a_{11} b_{10}} |1\rangle \; .
\end{equation}
It follows
\begin{equation} \label{line15}
\theta' \psi_1' = \psi_2' , \quad
\theta' \psi_2' = \psi_1'        \; .
\end{equation}
One further computes
\begin{equation} \label{line16}
A \psi_{1,2} = \sqrt{a_{00} a_{11}} \psi_{1,2}' , \quad
B \psi_{1,2} = \sqrt{b_{01} b_{10}} \psi_{1,2}'
\end{equation}
and
\begin{displaymath}
A |\psi_j \rangle\langle \psi_k| A^* =
|a_{00} a_{11}| |\psi_j' \rangle\langle \psi_k'| \; ,
\end{displaymath}
\begin{displaymath}
B |\psi_j \rangle\langle \psi_k| B^* =
|b_{01} b_{10}| |\psi_j' \rangle\langle \psi_k'| \; ,
\end{displaymath}
From here we can find easily an explicit expression for
the positive definite 2x2-matrix $\{r_{jk}\}$ fulfilling
\begin{displaymath}
\Phi(|\psi_j \rangle\langle \psi_k|) =
r_{jk} \; |\psi_j' \rangle\langle \psi_k'| \; .
\end{displaymath}
\medskip

{\bf 6.} \, Lemma 5 can be reformulated in the following way.
Every $\omega$ is within a certain line segment of constant
concurrence and constant $E(\Phi, .)$. That implies that
$\omega$ in
$E(\Phi, \omega)$ can be replaced by a pure state $\pi$ such
that $\omega$ is a convex combination of the optimal pair
$\pi$, $\theta \pi \theta$ without changing $E$, i.e.
\begin{displaymath}
E(\Phi, \omega) =
E(\Phi, \pi) = E(\Phi, \theta \pi \theta)
= S(\Phi(\pi)) = S(\Phi(\theta \pi \theta)) \; .
\end{displaymath}
If $\omega = \theta \omega \theta$ then we choose
\begin{displaymath}
\omega = \frac{1}{2} (\pi + \theta \pi \theta) \; .
\end{displaymath}
Varying $\pi$, we obtain a sufficient set to maximize
Holevo's quantity:
\begin{equation} \label{holevo4}
\chi^*(\Phi) = \max_{\pi} \;
S(\; \Phi(\frac{\pi + \theta \pi \theta}{2}) \;)
- S(\Phi(\pi)) \; ,
\end{equation}
the maximum is running through all pure states $\pi$.
A more symmetric looking, but identical expression reads
\begin{displaymath}
\chi^*(\Phi) = \max_{\pi} \;
S(\; \Phi(\frac{\pi + \theta \pi \theta}{2}) \;)
- \frac{S(\Phi(\pi)) + S(\Phi(\theta \pi\theta))}{2}
\end{displaymath}
By (\ref{line13}) we have
\begin{displaymath}
\Phi(\frac{\pi + \theta \pi \theta}{2}) =
\frac{\Phi(\pi) + \theta' \Phi(\pi) \theta'}{2}
\end{displaymath}
and, therefore,
\begin{equation} \label{holevo5}
\chi^*(\Phi) = \max_{\pi} \;
S(\; \frac{\Phi(\pi) + \theta' \Phi(\pi) \theta'}{2} \;)
- \frac{S(\Phi(\pi)) + S(\theta' \Phi(\pi) \theta')}{2}
\end{equation}
which is equivalent also to
\begin{displaymath}
\chi^*(\Phi) = \max_{\pi} \;
S(\; \frac{\Phi(\pi) + \theta' \Phi(\pi) \theta'}{2} \;)
- S(\Phi(\pi)) \; .
\end{displaymath}
Remarkably, we get the mean of a completely positive and
a completely co-positive map as argument of the entropy
function.

\end{document}